\documentclass[12pt,a4paper]{article}

\usepackage{latexsym}
\usepackage{amsmath}
\usepackage{amsfonts}
\usepackage{amssymb}
\usepackage{amsthm}
\usepackage{amscd}
\usepackage{mathrsfs}

\newcommand{\be}{\begin{equation}}
\newcommand{\ee}{\end{equation}}
\newcommand{\bea}{\begin{eqnarray}}
\newcommand{\eea}{\end{eqnarray}}

\def\no{\noindent}                      %
\def\cP{{\cal P}_y}                      %
\def\cG{{\cal G}}                       %
\def\G{{\cal G}}                        %
\def\cA{{\cal A}}                       %
\def\ri{{\mathrm{i}}}                   %
\def\cR{{\cal R}}                       %
\def\cJ{{\cal J}}                       %
\def\bR{{\mathbb R}}                    %
\def\bC{{\mathbb C}}                    %
\def\bZ{{\mathbb Z}}                    %
\def\bT{{\mathbb T}}                    %
\def\1{{\mbox{\boldmath $1$}}}          %
\def\tr{\mathrm{tr\,}}                  %
\def\cF{{\cal F}}                       %
\def\bt{\beta}                          %
\def\dt{\delta}                         %
\def\jp{\frac{1}{2}}                    %
\def\om{\omega}                         %
\def\Om{\Omega}                         %
\def\omfs{ \omega_{\mathrm{FS}}}        %
\def\al{\alpha}                         %
\def\ga{\gamma}                         %
\def\hb{\hat\beta}                      %
\def\ha{\hat\alpha}                     %
\def\val{\vert  y \vert}                %
\def\cp{\bC P(n-1)}                     %
\def\cE{{\cal E}}                       %
\def\cD{{\cal D}}                       %
\def\Hb{H^\mathrm{loc}}                 %
\begin{document}

\vspace*{0.5cm}
\begin{center}
{\Large \bf  On the spectra of the quantized action-variables of the
compactified Ruijsenaars-Schneider system} \end{center}

\vspace{0.2cm}

\begin{center}
L. Feh\'er${}^{a}$ and C. Klim\v c\'\i k${}^b$  \\

\bigskip

${}^a$Department of Theoretical Physics, WIGNER RCP, RMKI\\
H-1525 Budapest, P.O.B.~49,  Hungary, and\\
Department of Theoretical Physics, University of Szeged\\
Tisza Lajos krt 84-86, H-6720 Szeged, Hungary\\
e-mail: lfeher@rmki.kfki.hu

\bigskip

${}^b$Institut de math\'ematiques de Luminy
 \\ 163, Avenue de Luminy \\ F-13288 Marseille, France\\
 e-mail: klimcik@iml.univ-mrs.fr

\bigskip

\end{center}

\vspace{0.2cm}

\begin{abstract} A simple derivation of the spectra of the   action-variables of
the quantized compactified Ruijsenaars-Schneider system is presented.
The spectra are obtained by combining K\"ahler quantization with
the identification of the classical action-variables
as a standard toric moment map on the complex projective space.
The result is consistent with the Schr\"odinger quantization
of the system worked out previously
by van Diejen and Vinet.

\end{abstract}

\newpage
\section{Introduction}
\setcounter{equation}{0}

The definition of the compactified Ruijsenaars-Schneider system \cite{RIMS95} begins with the local
Hamiltonian\footnote{The index $k$ in the next product $\prod_{k\neq j}^n$ runs over
$\{1,2,...,n\}\setminus \{j\}$, and similar notation is used throughout.}
\be
\Hb(x, p) \equiv \sum_{j=1}^n \cos p_j \prod_{k\neq j}^n \left[1 - \frac{\sin^2 y}{
\sin^2 \frac{a(x_j-x_k)}{2}} \right]^{\frac{1}{2}}.
\label{I.1}\ee
The variables $\delta_j= e^{\ri a x_j}$ $(j=1,...,n)$
are interpreted as the positions
of $n$ ``particles'' moving on the circle and the canonically conjugate
 momenta $p_j$ encode the compact
variables  $\Theta_j = e^{-\ri p_j}$.
The parameters  $a$ and $y$  play the role of coupling constants.
Here,  the
center of mass condition $\prod_{j=1}^n \delta_j = \prod_{j=1}^n \Theta_j =1$ is also adopted.
Denoting the standard maximal torus of $SU(n)$ as $S\bT_n$, the
local phase space is
\be
M_y^{\mathrm{loc}}\equiv  \{ (\delta, \Theta)\,\vert\,
\delta= (\delta_1,...,\delta_n)\in  \cD_y,\, \Theta = (\Theta_1,...,\Theta_n)\in  S\bT_n\},
\label{I.2}\ee
where the domain $\cD_y \subset S\bT_n$ is chosen in such a way
to guarantee that $H^{\mathrm{loc}}$ takes real values.
The non-emptiness of $\cD_y$ is ensured by the restriction
$ \vert y \vert < \frac{\pi}{n}$.
The symplectic form on $M_{y}^{\mathrm{loc}}$ is defined by
\be
\Omega^{\mathrm{loc}}_a\equiv \frac{1}{a} \tr\!\left( \delta^{-1}
d\delta \wedge \Theta^{-1} d\Theta\right)
= \sum_{j=1}^n  dx_j \wedge dp_j  .
\label{I.3}\ee
The Hamiltonian $H^{\mathrm{loc}}$ can be recast as
the real part of the trace of the unitary Lax matrix $L_{y}^{\mathrm{loc}}$:
\be
L_{y}^{\mathrm{loc}}(\delta,\Theta)_{jl}\equiv
\frac{e^{\ri y} - e^{-\ri y}}{e^{\ri y}\delta_j \delta_l^{-1} - e^{-\ri y} }
W_j(\delta,y) W_l(\delta,-y) \Theta_l
\label{I.4}\ee
with the positive functions
\be
W_j(\delta,y):= \prod_{k\neq j}^n  \left[ \frac{ e^{\ri y} \delta_j  -e^{-\ri y} \delta_k }
{\delta_j - \delta_k}  \right]^{\frac{1}{2}}.
\label{I.5}\ee
The flows generated by the spectral invariants of $L_{y}^{\mathrm{loc}}$ commute, but are not
complete on $M_{y}^{\mathrm{loc}}$.
Ruijsenaars \cite{RIMS95} has shown that one can realize
$(M_{y}^{\mathrm{loc}}, \Omega^{\mathrm{loc}}_a)$ as a dense open submanifold
of the complex projective space $\bC P(n-1)$ equipped with a multiple
of the  Fubini-Study symplectic
form, and thereby the commuting local flows generated by
$L_{y}^{\mathrm{loc}}$
 extend to complete Hamiltonian flows on the compact phase space $\bC P(n-1)$.
 Ruijsenaars himself referred to
 the extended system on $\bC P(n-1)$  with complete flows as to the
 compactified  $\mathrm{III}_\mathrm{b}$ system.

 A rather complete solution of the quantized
 compactified Ruijsenaars-Schneider system was obtained  by van Diejen and Vinet \cite{vDV}
 by means of  ``Schr\"odinger quantization'', that is  by explicit diagonalization
 of the commuting Hamiltonian operators in a coordinate representation.
 Here, our main purpose is to give an alternative, very simple, derivation of
 the result of \cite{vDV} regarding the joint spectrum of the action-variables.
We note that our parameter $a>0$ corresponds to
$\alpha$ in \cite{vDV}
and the constants  $g,M$ that appear
in \cite{vDV} are related to our parameters $y,a$ above by
\be
g\equiv \frac{2 \vert y\vert}{a},
\quad
M\equiv\frac{2}{a} (\pi - n \vert y\vert).
\label{I.6}\ee
We follow the convention of \cite{FK} in the present paper, except that in this reference
the constant $a$ was set equal to $2$.

Classically,  the action-variables can be identified with the components of the moment map
of a standard Hamiltonian $\bT_{n-1}$ torus action on $(\cp, M \omega_{\mathrm{FS}})$,
and we shall obtain the quantized spectra by combining this identification
with the K\"ahler quantization of the projective space.
The identification just mentioned relies on a very non-trivial symplectic automorphism of $\cp$
that encodes the self-duality of the system discovered by Ruijsenaars \cite{RIMS95}.
In  our recent paper \cite{FK}, we found a geometric interpretation
of the self-duality.
Namely, we have shown that the compactified Ruijsenaars-Schneider system can be
 obtained by an appropriate quasi-Hamiltonian reduction of the internally fused
  quasi-Hamiltonian double\footnote{Via the connection between Chern-Simons models
  and quasi-Hamiltonian geometry \cite{AMM},
this confirms the related conjectures of Gorsky and Nekrasov \cite{GN}.}
 $SU(n)\times SU(n)$, and the self-duality is due to the  democracy enjoyed by the two
 $SU(n)$ factors of the double.
 More specifically, we  demonstrated  that   the reduced phase space is a Hamiltonian
 toric manifold and that its
 identification with $\bC P(n-1)$ follows  from the Delzant theorem of  symplectic topology.
Next we  recall the notion of quasi-Hamiltonian reduction
 \cite{AMM} and then review the pertinent results of \cite{FK}. Finally we present the
derivation of the quantized spectra, which is an easy by-product of the geometric picture.

 \section{Quasi-Hamiltonian reduction}
 \setcounter{equation}{0}

 Consider a Lie group $G$ acting  on a  manifold $D$.
For a  $G$-invariant \emph{symplectic} form $\omega$ on  $D$ the corresponding Poisson
bracket $\{\alpha,\beta\}$ of
$G$-invariant functions $\al,\beta$  is again $G$-invariant.  The quotient $D/G$
 is thus
naturally equipped with a Poisson structure. The choice of a
symplectic leaf of this Poisson structure on
$D/G$ is known as a symplectic reduction of $D$.

Alekseev, Malkin and Meinrenken have shown that a  $G$-invariant
 $2$-form $\omega$  on a manifold $D$ obeying the axioms of quasi-Hamiltonian
geometry  \emph{need not be symplectic}    and still induces
a Poisson structure on $D/G$ \cite{AMM}. This happens because the
axioms of quasi-Hamiltonian
geometry imply that  for each $G$-invariant function $\alpha$ there exists an unique
$G$-invariant vector field $v_\alpha$ on $D$  verifying
 \be
\omega( v_\alpha, \cdot)=d\alpha,
\label{(*)}\ee  and, moreover,  the function
\be\{\alpha,\beta\}:=\omega(v_\alpha,v_\beta)\label{(**)}\ee is $G$-invariant and it
defines
 a Poisson bracket on the space of the  $G$-invariant functions on $D$.
The choice of a symplectic leaf of this Poisson structure on $D/G$ is called a
 quasi-Hamiltonian reduction of $D$.

 Of course $D/G$ is not a smooth manifold in general, but nevertheless the above remarks
convey the main idea of reduction.
Below we recall the axioms of quasi-Hamiltonian geometry
and the reduction procedure  in a more precise manner.

Let $G$ be a compact Lie group with Lie algebra $\G$.
Fix an invariant scalar product
$\langle\cdot,\cdot\rangle$ on $\G$ and denote by $\vartheta$ and $\bar\vartheta$,
   respectively, the left- and right-invariant Maurer-Cartan forms on $G$.
   For a  $G$-manifold $D$ with action $\Psi: G\times D\to D$, we
   use $\Psi_\eta(x):= \Psi(\eta,x)$ and
   let $\zeta_D$ denote the vector field on $D$
   that corresponds to $\zeta\in \G$.
   The adjoint action of $G$ on itself is given by $\mathrm{Ad}_\eta (\tilde\eta):=
   \eta \tilde\eta \eta^{-1}$,
   and $\mathrm{Ad}_\eta$ denotes also the induced action on $\G$.

   By definition \cite{AMM}, a quasi-Hamiltonian $G$-space $(D,G,\om,\mu)$ is a $G$-manifold
   $D$  equipped
   with an invariant $2$-form $\om\in\Lambda^2(D)^G$ and with an equivariant  map $\mu:D\to G$,
$\mu \circ \Psi_\eta = \mathrm{Ad}_\eta \circ \mu$,
 in such way that the following conditions hold.

   \noindent (a1) The differential of $\omega$ is given by
\be
   d\omega =-\frac{1}{12} \mu^*\langle\vartheta,[\vartheta,\vartheta]\rangle.
\ee
\noindent (a2) The infinitesimal action is related to $\mu$ and $\omega$  by
\be
\om(\zeta_D,\cdot) = \jp\mu^*\langle\vartheta+\bar\vartheta,\zeta\rangle,
\quad \forall \zeta\in\G.
\ee
\noindent (a3) At each $x\in D$, the kernel of $\om_x$ is provided by
\be
   \mathrm{Ker}(\om_x)=\{\zeta_D(x)\,\vert\,
   \zeta\in \mathrm{Ker}(\mathrm{Ad}_{\mu(x)}+ \operatorname{Id}_\cG)\}.
\ee
The map $\mu$ is called the moment map.

Any $G$-invariant function $h$ on $D$ induces a unique vector field
$v_h$ on $D$ that satisfies
(\ref{(*)})  and  preserves
 $\mu$ as well as $\omega$.
A quasi-Hamiltonian dynamical system
$(D,G,\om,\mu,h)$ is a quasi-Hamiltonian
   $G$-space with a distinguished
   $G$-invariant function $h\in C^\infty(D)^G$, the Hamiltonian.

The quasi-Hamiltonian reduction of a quasi-Hamiltonian dynamical system  $(D,G,\om,\mu,h)$ can
 be determined by choosing an element $\mu_0\in G$. We say that
    $\mu_0$ is {\it strongly regular} if it satisfies the following two conditions:
\begin{enumerate}
 \item{
 The subset  $\mu^{-1}(\mu_0):=\{   x\in D\,\vert\,  \mu(x)=\mu_0\}$  is an embedded
 submanifold of $D$.}

 \item{
   If  $G_0 < G$ is  the isotropy group of $\mu_0$ with respect to the adjoint action, then the
   quotient  $\mu^{-1}(\mu_0)/G_0$ is a manifold for which the canonical projection
$p:\mu^{-1}(\mu_0)\to \mu^{-1}(\mu_0)/G_0$ is a smooth submersion.}
\end{enumerate}
The  result of the reduction based on a strongly regular element $\mu_0$  is
 a standard Hamiltonian system,  $(P,\hat\om,\hat h)$.
The reduced phase space $P$ is the manifold
\be
P \equiv \mu^{-1}(\mu_0)/G_0,
\label{2.6}\ee
which carries
the reduced symplectic form $\hat\om$ and reduced Hamiltonian $\hat h$ uniquely defined by
\be
p^*\hat\om=\iota^*\om, \quad p^*\hat h=\iota^*h,
\label{genred}\ee
where $\iota:\mu^{-1}(\mu_0)\to D$ is the tautological embedding.

We stress that $\hat \omega$ is a symplectic form in the usual sense, whilst $\omega$
is neither  closed
nor globally non-degenerate in general.
The Hamiltonian vector field and the flow defined by
$\hat h$ on $P$ can be obtained by first restricting
the quasi-Hamiltonian
vector field $v_h$ and its flow to the ``constraint surface'' $\mu^{-1}(\mu_0)$ and then applying
the canonical projection $p$.

We wish to note for clarity  that the group valued
quasi-Hamiltonian moment map utilized here
is different from the group valued Poisson-Lie moment map introduced
by Lu \cite{Lu}. We used the Poisson-Lie
moment map in a previous paper \cite{FKinCMP} to describe the standard
trigonometric Ruijsenaars-Schneider system and its dual \cite{RIMS95},
which are different from the system (\ref{I.1}).

\section{The reduction of the internally fused double}
\setcounter{equation}{0}

 Let $G\equiv SU(n)$.
  The quasi-Hamiltonian manifold $(D,G,\omega,\mu)$ that we are going to reduce is provided
   by the Cartesian product
{\be
  D:=  G \times  G =\{ (A,B)\,\vert\, A,B\in   G\}.
\label{2.1}\ee}
The invariant scalar product on Lie($G$)$=  su(n)$ is given by
\be
\langle \zeta, \tilde\zeta\rangle := -\frac{1}{a} \tr(\zeta \tilde \zeta),
\qquad
\forall \zeta, \tilde\zeta  \in su(n).
\label{scalarprod}\ee
The group $G$ acts  on $D$ by componentwise conjugation
{ \be
\Psi_\eta(A,B):=  (\eta A  \eta^{-1}, \eta B \eta^{-1}).
\label{2conj}\ee}
The  $2$-form $\omega$ on  $D$  reads
 \be\omega:= \frac{1}{a} \langle A^{-1} dA \stackrel{\wedge}{,} dB B^{-1}\rangle
+ \frac{1}{a}  \langle  dA A^{-1} \stackrel{\wedge}{,} B^{-1} dB \rangle
-\frac{1}{a} \langle (AB)^{-1} d (A B) \stackrel{\wedge}{,} (BA)^{-1} d (BA) \rangle,
\label{2.2}\ee
 and the $G$-valued moment map $\mu$  is defined by
  \be
\mu(A,B)= AB A^{-1} B^{-1}.
\label{1.1}\ee
In order to define a quasi-Hamiltonian dynamical system on the
``internally fused double'' $D$ described above \cite{AMM} we need also a $G$-invariant Hamiltonian.
We shall in fact consider two families of such Hamiltonians each one containing $(n-1)$ members.
They are given by the so-called
\emph{spectral functions} on  each $SU(n)$  factor of  the double.

Roughly speaking, the spectral functions on $SU(n)$ evaluated at a point $C\in SU(n)$
are defined as  logarithms of  ratios of two neighbouring eigenvalues of $C$. More precisely,
we define the alcove $\cA$  by
\be
\cA:=\Bigl\{ (\xi_1,..., \xi_n)\in \bR^n\,\Big\vert\, \xi_j \geq 0,
\quad j=1,...,n, \quad \sum_{j=1}^n \xi_j=\pi\Bigr\}
 \ee
 and consider the injective map $\dt$ from $\cA$ into the
 subgroup $S\bT_n$ of  the diagonal elements
 of $SU(n)$  given by
\be \dt_{11}(\xi):=e^{\frac{2\ri}{n}\sum_{j=1}^n j\xi_j},\quad
\dt_{kk}(\xi):=e^{2\ri \sum_{j=1}^{k-1}\xi_j}\dt_{11}(\xi),
 \quad k=2,...,n.
 \label{par1}\ee
With the aid of the fundamental weights $\Lambda_k$ of $su(n)$
represented by the diagonal matrices
$\Lambda_k \equiv \sum_{j=1}^k E_{jj} - \frac{k}{n} \1_n$,
the matrix $\delta(\xi)$ can be written in the form
\be
\delta(\xi) = \exp\left(-2 \ri \sum_{k=1}^{n-1} \xi_k \Lambda_k\right).
\label{3.9}\ee
Every element $C\in SU(n)$ can be diagonalized as
\be
C=\eta^{-1}\dt(\xi)\eta, \label{D}
 \ee
for some $\eta\in SU(n)$ and { unique}  $\xi\in \cA$.
By definition, the $j^{\,\mathrm{th}}$ component $\xi_j$ ($j=1,...,n$) of the alcove element $\xi$
entering the decomposition (\ref{D}) {  is
the value of  the spectral function} $\xi_j$ on $C\in SU(n)$.

Consider  now
the  $2(n-1)$ distinguished $G$-invariant  Hamiltonians  $\al_j$, $\bt_j$ on the double $D$
defined in terms of the spectral functions according to
\be
\al_j(A,B):=\frac{2}{a} \xi_j(A),\quad
\bt_j(A,B):=\frac{2}{a}\xi_j(B), \quad j=1,...,n-1.
\label{3.11}\ee
It turns out that for each $j$ the  associated quasi-Hamiltonian
 vector fields $v_{\al_j}$    can be integrated
 to a  very simple {  circle  action}  on $D$:
 \be
 \left(A,B\eta(A)^{-1}\mathrm{diag}(1,1,...,1,e^{\ri t},e^{-\ri t},1,...,1)\eta(A)\right),
 \quad t\in\bR.
 \label{3.12}\ee
 Here  the phase $e^{\ri t}$ sits in the $j^{\,\mathrm{th}}$ entry of the diagonal and $\eta(A)$
 is given by
 the diagonalization  $A=\eta(A)^{-1}\dt(\xi)\eta(A)$.
Similarly,  the following   $2\pi$-periodic curve  in $D$ is an integral curve of
the vector field $v_{\bt_j}$:
 \be
 \left(A\eta(B)^{-1}{\rm diag}(1,1,...,1,e^{-\ri t},e^{\ri t},1,...,1)\eta(B),B\right), \quad
 t\in\bR.
 \label{3.13}\ee
When taken together for every $j=1,...,n-1$, the flows (\ref{3.12})
give rise to the $\alpha$-generated torus action, and the flows (\ref{3.13})
yield the $\beta$-generated torus action on $D$.
We note in passing that the spectral Hamiltonians $\alpha_j$ and $\beta_j$ are smooth
only on the dense open subset $D_{\mathrm{reg}} \subset D$ where $A$ and
$B$ have distinct eigenvalues, but this does not
lead to any difficulty since the constraint surface of our reduction turns out to be a submanifold
of $D_{\mathrm{reg}}$.

Now we  state two  theorems proved in \cite{FK} that characterize
  our reduction
 of the quasi-Hamiltonian dynamical systems based on the spectral Hamiltonians
 $\al_j$ and $\beta_j$ on the double.

\medskip
\noindent
{\bf Theorem  1.}  \emph{The choice $\mu_0 = {\rm diag}(e^{2\ri y},...,e^{2\ri y},e^{2(1-n)\ri y})$
defines a  strongly regular value of the moment map $\mu$  (\ref{1.1})  whenever the
real parameter $y$ verifies the condition
$0<\vert y\vert <\frac{\pi}{n}$.
The corresponding  reduced phase space $P $ (\ref{2.6}) is a smooth,
compact manifold of dimension $2(n-1)$.}

   \medskip\no
   {\bf Theorem 2.}  \emph{The reduced phase space  $P =\mu^{-1}(\mu_0)/G_0$ is connected.
   Moreover, both the $(\al_1,...,\al_{n-1})$-generated
   and   the $(\bt_1,...,\bt_{n-1})$-generated torus actions on the
double descend to the reduced phase space $P$,
    where they become Hamiltonian and effective.}

\medskip
 Let us recall  that
   a \emph{Hamiltonian toric manifold}\footnote{A review of these compact completely
   integrable systems
 can be found in \cite{Cannas}.}
   is a compact connected symplectic manifold of
   dimension $2(n-1)$  equipped with an effective  Hamiltonian action of a torus of dimension
   $(n-1)$. {Theorems 1 and 2 ensure that
   the reduced phase space $(P, \hat \omega)$ (\ref{genred})
    is a Hamiltonian toric manifold} { in two different ways (i.e. $\al$-generated way
    and $\beta$-generated way).

The next theorem, proved again  in \cite{FK},  gives the key for the identification
of the reduced phase space with the complex projective
space $\bC P(n-1)$.

\medskip

\no  {\bf Theorem 3.} \emph{The common  image of the reduced phase space $P$ under
both $(n-1)$-tuples of  reduced
  spectral Hamiltonians
 $(\ha_1,...,\ha_{n-1})$ and $(\hb_1,...,\hb_{n-1})$  is the convex polytope
  $\frac{2}{a}\cP$, where
 \be
 \cP:=\Bigl\{(\xi_1,...,\xi_{n-1})\in \bR^{n-1}\,\Big\vert\,
  \xi_j \geq \vert y \vert, \,\,\, j=1,...,n-1, \,\,\,
  \sum_{j=1}^{n-1} \xi_j\leq \pi-\val\Bigr\}.
  \label{3.14}\ee}
\medskip

In order to be able to put Theorem 3 to use, and also for reference in Section 4,
we need to sketch an auxiliary  symplectic reduction treatment of $\cp$.
For this, take the symplectic vector space $\bC^n \simeq \bR^{2n}$ endowed with the Darboux form
$\Om = \ri \sum_{k=1}^n d\bar u_k \wedge du_k$,
where $u_k$ are the components of  the vector $u$ that runs over $\bC^n$.
Then consider the Hamiltonian action $\psi$ of the group $U(1)$ on $\bC^n$ operating as
$\psi_{e^{\ri \rho}}(u):= e^{\ri \rho} u$.
This $U(1)$ action is generated by the moment map $\chi: \bC^n \to \bR$,
\be
\chi(u)\equiv \sum_{k=1}^n J_k
\quad\hbox{with}\quad
J_k:= \vert u_k \vert^2 \qquad (\forall k=1,...,n).
\label{3.15}\ee
For any fixed value $M>0$,
 ordinary (Marsden-Weinstein)  symplectic reduction of $(\bC^n , \Om)$ yields the reduced phase space
\be
\chi^{-1}(M)/U(1) \equiv \bC P(n-1).
\label{3.16}\ee
The corresponding reduced symplectic form is $M\omfs$, where
$\omfs$ is the standard Fubini-Study form  of $\bC P(n-1)$.
The functions $J_k$ are $U(1)$ invariant and thus descend to smooth functions
on the reduced phase space $(\cp, M\omfs)$, which we shall denote below as $\hat J_k$.

Now focus on the action
$R: \bT_{n-1} \times \bC^n \to \bC^n$
of the torus $\bT_{n-1}$ on $\bC^n$  furnished by
\be
R_\tau (u_1, ..., u_{n-1}, u_n) := (\tau_1 u_1, ..., \tau_{n-1}u_{n-1}, u_n),
\quad
\forall \tau\in \bT_{n-1},\,\, \forall u\in \bC^n.
\label{Ract}\ee
The corresponding moment map can be taken to be $J= (J_1, ..., J_{n-1}) : \bC^n \to \bR^{n-1}$.
Of course, the toric moment map is unique only
up to a shift by an arbitrary constant.

The  $\bT_{n-1}$-action (\ref{Ract}) and its moment map survive the symplectic reduction
by the $U(1)$-action $\psi$  and give rise to the so-called
``rotational $\bT_{n-1}$-action'' on $(\bC P(n-1), M \omega_{\mathrm{FS}})$,
which thus becomes a Hamiltonian toric manifold.
The rotational $\bT_{n-1}$-action $\cR: \bT_{n-1} \times \bC P(n-1)$
operates according to the rule
$\cR_\tau \circ \pi_{M} = \pi_{M} \circ R_\tau$,
where $\pi_{M}: \chi^{-1}(M)\to \bC P(n-1)$ is the canonical projection.
We choose its moment map to be
\be
 \cJ= (\cJ_1, ..., \cJ_{n-1}): \bC P(n-1) \to \bR^{n-1}
 \quad
 \hbox{with}
\quad
\cJ_k  := \hat J_k + g.
\ee
Here the constant $g$ is related to  $y$  and $a$ as given previously (\ref{I.6}).
It is then easily seen that \emph{the
image of $\cp$ under the toric moment map $\cJ$ is the same polytope
$\frac{2}{a}\cP$ that features in our Theorem 3.}
 We can therefore use the following celebrated result.

\medskip
 \no {\bf Delzant's theorem \cite{De}.}
 \emph{Let $(M_1, \omega_1,\Phi_1)$ and $(M_2, \omega_2,\Phi_2)$
  be  Hamiltonian toric manifolds.  If the images of the  moment maps
  $\Phi_1(M_1)$ and $\Phi_2(M_2)$ coincide,
then there exists a torus-equivariant symplectomorphism $\phi: M_1 \to M_2$
for which $\Phi_1 = \Phi_2 \circ \phi$.}

\medskip

The combination of the above statements directly leads to
one of the main results of \cite{FK}.

\medskip
\no
{\bf Corollary.} \emph{All three Hamiltonian toric manifolds
$(\bC P(n-1), M \omega_{\mathrm{FS}},  \cJ)$,
$(P,\hat\om, \hat \alpha)$ and $(P, \hat \om, \hat \beta)$
are equivariantly symplectomorphic to each other.}

\medskip

By applying the corollary, we have
``Delzant symplectomorphisms'' $\varphi_\al,\varphi_\bt: \bC P(n-1)\to P$
that are subject to
\be
\varphi_\alpha^*\hat \omega=   \varphi_\beta^*\hat \omega= M  \omega_{\mathrm{FS}},
\qquad\varphi_\alpha^*\ha=   \varphi_\beta^*\hb=\cJ.
\label{Delz}\ee
In fact, the symplectomorphisms $\varphi_\beta$ and $\varphi_\alpha$ give rise to
two models of $(P, \hat\om, \ha, \hb)$ in such a  way that in terms of model (i)
the functions $\hb_k$ become the particle-positions and the functions $\ha_k$ become
the action-variables of the compactified Ruijsenaars-Schneider system,
and their role is interchanged  in model (ii).
In both cases the model of $(P,\hat\om)$ itself is provided by $(\cp, M \omfs)$,
which serves as the completed phase space of the  $\mathrm{III}_\mathrm{b}$ system
equipped with the global particle-position variables  $\cJ_k$.
Before explaining these statements, we need some more preparation.

First, following \cite{RIMS95,FK}, we introduce Darboux coordinates
on the dense open submanifold $\cp_0 \subset \cp$ where none of
the homogeneous coordinates can vanish.
To do this, consider the manifold
\be
\frac{2}{a} \cP^0  \times \bT_{n-1} = \{ (\gamma, \tau)\}.
\ee
Here $\cP^0$ is the interior of the polytope (\ref{3.14}) and we also write
$\tau_j= e^{\theta_j}$ ($j=1,...,n-1)$.
We define the map $\cE: \frac{2}{a} \cP^0 \times \bT_{n-1} \to \chi^{-1}(M)$ by
requiring that
$\cE: (\gamma, \tau) \mapsto  (u_1,...,u_{n-1}, u_n)$  according to
\be
u_j= \tau_j \sqrt{\gamma_j - g}\quad (j=1,...,n-1),
\qquad u_n=\sqrt{M + (n-1) g - \sum_{j=1}^{n-1} \gamma_j}.
\label{embed}\ee
Then  $\pi_M \circ \cE: \frac{2}{a} \cP^0  \times \bT_{n-1} \to \cp_0$ is a
diffeomorphism satisfying
\be
(\pi_M \circ \cE)^* (M \omfs)=
\ri \sum_{k=1}^{n-1} d \ga_k \wedge
d \tau_k \tau_k^{-1} = \sum_{k=1}^{n-1}  d  \theta_k \wedge d\ga_k.
\ee
Second, we identify the local phase space $M_y^{\mathrm{loc}} = \cD_y \times S\bT_n$ (\ref{I.2})
of the $\mathrm{III}_\mathrm{b}$ system with $\frac{2}{a} \cP^0  \times \bT_{n-1}$
by means of the map $\cF:(\gamma, \tau) \mapsto (\delta (a \gamma/2), \Theta(\tau))$
given by
\be
\delta (a \gamma/2)
= e^{-\ri a \sum_{k=1}^{n-1} \gamma_k \Lambda_k},
\qquad
\Theta(\tau) := e^{-\ri \sum_{k=1}^{n-1} \theta_k (E_{k,k} - E_{k+1, k+1})},
\ee
where the same notations is used for $\delta$ as in (\ref{3.9}).
The map $\cF$  verifies $\cF^* \Om^{\mathrm{loc}}_a =\sum_{k=1}^{n-1}  d  \theta_k \wedge d\ga_k$
and its formula shows that $\gamma$ represents the particle-positions
(or rather $(-1)$-times the particle positions since
$\delta_j = e^{\ri a x_j}$ was used in the Introduction)
of the
local $\mathrm{III}_\mathrm{b}$ system.
By combining the symplectic diffeomorphisms
\be
M_y^{\mathrm{loc}} \simeq \frac{2}{a} \cP^0  \times \bT_{n-1}\simeq \cp_0,
\ee
we can identify $M_y^{\mathrm{loc}}$ with $\cp_0$.
Thereby the particle-positions, $\gamma_k$, turn into the rotational moment
map, $\cJ_k = \vert u_k \vert^2 + g$, that remains well-defined on the whole of $\cp$.

Now we return to the Delzant symplectomorphisms, and quote the following result from
\cite{FK}.

\medskip

\no  {\bf Theorem 4.}
\emph{The Delzant symplectomorphism $\varphi_\beta$ (\ref{Delz}) can be chosen so that
its restriction to $\cp_0$ operates according to the
following explicit formula:
\be
\varphi_\beta \left(\pi_M\circ \cE(\gamma,\tau)\right) =
 p \circ \Psi_{\eta^{-1}}
 \left( L_{y}^{\mathrm{loc}}(\delta(a\gamma/2),\Theta(\tau)),
 \delta(a\gamma/2) \right),
 \label{3.21}\ee
 where $\eta$ is any $U(n)$ matrix the last column of which is
proportional to the vector
\be
v_j(a\gamma/2,y):=
\left[\frac{\sin y}{\sin ny}\right]^{\frac{1}{2}} W_j(\delta(a\ga/2), y),
\qquad j=1,...,n.
\label{vW}\ee
Here the notations (\ref{I.4})-(\ref{I.5}) are used, and $\Psi_{\eta^{-1}}$ acts by
componentwise conjugation  (\ref{2conj}).
The ambiguity in the definition of $\eta\in U(n)$ is killed by the
projection map $p:\mu^{-1}(\mu_0)\to P$.}

\medskip
One of the non-trivial points of Theorem 4 is that
the map given on $\cp_0$ by (\ref{3.21}) extends to
a \emph{globally well-defined}  symplectomorphism $\varphi_\beta: \cp \to P$.
We have not (yet) studied what is the most general global Delzant symplectomorphism,
since formula (\ref{3.21}) gives one and this is enough for our purpose.
Namely, it follows from formula (\ref{3.21}}) with (\ref{3.11})  that
\bea
&& \ha_k\circ \varphi_\beta \left(\pi_M\circ \cE(\gamma,\tau)\right) =
\frac{2}{a}\xi_k\!\left(L_{y}^{\mathrm{loc}}\left(\delta(a\gamma/2),\Theta(\tau)\right)\right),
\nonumber\\
&& \hb_k\circ \varphi_\beta \left(\pi_M\circ \cE(\gamma,\tau)\right)=
\cJ_k\left(\pi_M\circ \cE(\gamma,\tau)\right)= \gamma_k.
\label{3.26}\eea
This tells us that $\varphi_\beta$
\emph{converts
the reduced spectral Hamiltonians $\ha$ into the action-variables of the
compactified Ruijsenaars-Schneider system, whose global particle-position
variables on the completed phase space $\cp$ are furnished by the  function $\cJ$}.

In the paper \cite{FK} we also gave the analogous local formula of
 $\varphi_\alpha$.
It shows that the application of the  pull-back $\varphi_\alpha^*$
converts the functions $\ha_k$ into the global particle-positions
and the functions $\hb_k$ into the action-variables of the compactified Ruijsenaars-Schneider system.
Then it also follows that the symplectic automorphism
\be
\phi:= \varphi_\alpha^{-1} \circ \varphi_\beta: \cp \to \cp
\label{RSdual}\ee
is nothing but Ruijsenaars' self-duality map for the compactified system \cite{RIMS95}
that converts  the particle-positions into the action-variables, and vice versa.

We finish this section
with a remark clarifying  the relationship between the two Delzant symplectomorphims
$\varphi_\beta$ and $\varphi_\alpha$ and the involution property of the duality map
$\phi$ (\ref{RSdual}).
For this we need to note that the reduced phase space $P$ admits a natural anti-symplectic
involution, $\hat m$. In fact, $\hat m$  is induced from the anti-automorphism $m$ of the internally
fused double $D$ operating as $m(A,B):= (\bar B, \bar A)$,
where ``bar'' denotes complex conjugation.
Similarly, $\cp$ permits the anti-symplectic involution $\hat \Gamma$
induced by the anti-symplectic involution $\Gamma$ of the
symplectic vector space $(\bC^n, \Om)$,
$\Gamma(u_1, ..., u_{n-1}, u_n):= (\bar u_{n-1},..., \bar u_1, \bar u_n)$.
That is,  $\Gamma$ acts as  reflection composed with complex conjugation on the first $(n-1)$
coordinates, and  $u_n$ is chosen as special in accordance
with our embedding of $M_y^{\mathrm{loc}}$ into $\cp$ (cf.~(\ref{embed})).
Now, it can be shown that if $\varphi_\beta$ is any Delzant symplectomorphism verifying (\ref{Delz}),
then
\be
\varphi_\alpha := \hat m \circ \varphi_\beta \circ \hat \Gamma
\ee
also verifies the required properties. This implies the anti-symplectic
involution property
\be
(\hat \Gamma \circ \varphi_\alpha^{-1} \circ \varphi_\beta)^2= {\mathrm{id}}_{\bC P(n-1)}
\ee
and it can be also proved that $(\varphi_\alpha^{-1} \circ \varphi_\beta)^4 =
{\mathrm{id}}_{\bC P(n-1)}$ consistently with the results of \cite{RIMS95}.

 \section{Spectra from K\"ahler quantization}
\setcounter{equation}{0}

To sum up,
we saw in Section 3 that the reduced phase space $(P, \hat\omega)$
of the quasi-Hamiltonian reduction carries two  toric moment maps
defined by the two sets of functions $\hat \alpha_k$ and $\hat \beta_k$.
We then exhibited symplectomorphisms between $(P,\hat\om)$ and $(\cp,M\omfs)$
that bring either of these two moment maps into the standard
rotational toric moment map on $\cp$, yielding the correspondences
\be
 \hat \alpha_k \leftrightarrow (\hat J_k +g) \leftrightarrow
\hat \beta_k.
\label{4.1}\ee
 $\hat J_k \in C^\infty(\cp)$ descended from $J_k\in C^\infty(\bC^n)$ (\ref{3.15})
 by ordinary symplectic reduction.
According to (\ref{3.26}), $\hat \beta_k$ represent global analogues of the particle-positions
of the compactified Ruijsenaars-Schneider  system for which $\hat \alpha_k$
serve as the \emph{action-variables}.

Recall that the standard Ruijsenaars-Schneider Hamiltonians, ${\cal H}_r$ ($r=1,..., n-1$),
are the real parts of the
elementary symmetric functions of the
Lax matrix.
It follows from equations (\ref{3.21}), (\ref{3.26}) and  (\ref{3.9})-(\ref{3.11}) that
unique, globally smooth
extensions of
these Hamiltonians are furnished by the  elementary symmetric functions of
the matrix
\be
\delta(a\hat\alpha/2) = \exp\left(-\ri a \sum_{k=1}^{n-1}
\hat \alpha_k \Lambda_k\right).
\label{4.2}\ee
Hence, commuting quantum operators,
${\cal H}_r^{\mathrm{op}}$, can be defined by taking the
elementary symmetric functions of
\be
\delta(a \hat\alpha^{\mathrm{op}}/2) = \exp\left(-\ri a \sum_{k=1}^{n-1}
\hat \alpha_k^{\mathrm{op}} \Lambda_k\right),
\label{4.3}\ee
if one can construct commuting self-adjoint operators
$\hat \alpha_k^{\mathrm{op}}$  corresponding to
the action-variables.

Now we observe that the standard K\"ahler quantization of $(\cp, M \omfs)$
gives rise to a natural quantization of the functions $\hat J_k$, and through
the symplectomorphism
whereby $\hat \alpha_k \leftrightarrow (\hat J_k +g)$
 this can be used to quantize the variables $\hat \alpha_k$.
By this procedure,  \emph{the joint spectrum of
the resulting quantized action-variables follows immediately. }
Of course, application of the analogous  quantization procedure  to
$\hat \beta_k$ leads to  the same spectrum.

To develop the above observation,
let us outline
the  quantum mechanical counterpart of the classical reduction procedure
whereby we obtained $(\cp, M \omega_{\mathrm{FS}})$ from
the canonical symplectic space $(\bC^n,\Omega)$.
For this,  consider the standard holomorphic quantization of
$(\bC^n,\Omega)$. This engenders the  commuting self-adjoint operators
$J_l^{\mathrm{op}}= u_l \frac{\partial}{\partial u_l}$.
The joint orthonormal eigenvector basis of the operators
\be
( J_1^{\mathrm{op}},...,  J_n^{\mathrm{op}})
\label{4.4}\ee
can be described as the set of the
$n$-tuples
\be
\vert \nu_1,...., \nu_n \rangle
\qquad
\hbox{with any}\quad
\nu_l\in \bZ_{\geq 0},\quad
l=1,...,n.
\label{4.5}\ee
In other words,  $ J_l^{\mathrm{op}}$ $(l=1,...,n)$ are  ``number-operators''
of independent harmonic oscillators and
the $\nu_l$ are the corresponding eigenvalues.
As a holomorphic function
on the phase space $\bC^n$, up to a normalization constant,
$\vert \nu_1,...., \nu_n \rangle \sim \prod_{l=1}^n u_l^{\nu_l}$.

Now we can quantize $(\cp, M \omega_{\mathrm{FS}})$ by quantum Hamiltonian reduction.
In effect, this requires the imposition of  the quantum analogue of the
$U(1)$ moment map constraint,
\be
\sum_{k=1}^n J_k = M,
\label{4.6}\ee
on the states.
The reduced Hilbert space is spanned by
 those basis vectors $\vert \nu_1,...., \nu_n\rangle$ for which
\be
\sum_{k=1}^n \nu_k = M.
\label{4.7}\ee
Hence the \emph{quantization condition} $M\in \bZ_{>0}$ must be satisfied.
The resulting orthonormal basis of the reduced Hilbert space can be also
thought of as the set
of the $(n-1)$-tuples
\be
\vert \nu_1,...,\nu_{n-1}\rangle_{\mathrm{red}}
\quad
\hbox{subject to}\quad
\nu_k\in \bZ_{\geq 0},\quad
\sum_{k=1}^{n-1}\nu_k \leq M.
\label{redstate}\ee
Such an  $(n-1)$-tuple represents the state
\be
\vert \nu_1,..., \nu_{n-1}, M- \sum_{k=1}^{n-1} \nu_k \rangle
\label{4.10}\ee
in the original Hilbert space.
The procedure of quantum Hamiltonian reduction yields
the commuting  operators $\hat J_l^{\mathrm{op}}$ that act on
the reduced Hilbert space according to
\be
\hat J_l^{\mathrm{op}} \vert \nu_1,...,\nu_{n-1}\rangle_{\mathrm{red}}
= \nu_l \vert \nu_1,...,\nu_{n-1}\rangle_{\mathrm{red}}.
\label{4.11}\ee
In simplest terms, the operators $\hat J_l^{\mathrm{op}}$ are just the restrictions
of the operators $J_l^{\mathrm{op}}$ to the states verifying the constraint (4.7).

It is worth noting that
the reduced Hilbert space defined above is the same
as the outcome of the (geometric) K\"ahler quantization
of $(\cp, M \omega_{\mathrm{FS}})$.
This can be verified  by viewing the state (\ref{redstate}) as an
$U(1)$ equivariant complex function on  $\chi^{-1}(M)\subset \bC^n$ (cf.~(\ref{3.16})).
The function just alluded to is  the restriction of the function (\ref{4.10}), and
it encodes a holomorphic
section of the corresponding line bundle over $(\cp, M \omfs)$.
Incidentally,
every Hamiltonian toric manifold can be quantized in a similar way by
quantum Hamiltonian reduction, as expounded in a somewhat different language in \cite{Ham}.

By adopting the above K\"ahler quantization to
the action-variables of the compactified Ruijsenaars-Schneider system, we immediately obtain
from (4.1) that \emph{the
joint spectrum of the quantized action-operators
$\hat\alpha_k^{\mathrm{op}}$
is given by the $(n-1)$-tuples}
\be
(\nu_1+g,...,\nu_{n-1}+g)
\quad
\hbox{subject to}\quad
\nu_k\in \bZ_{\geq 0},\quad
\sum_{k=1}^{n-1}\nu_k \leq M.
\label{4.12}\ee
The joint spectrum
of the action-operators is non-degenerate (all joint eigenvalues  have multiplicity $1$),
and the eigenvalues  of
the commuting Hamiltonians  ${\cal H}_r^{\mathrm{op}}$ can be obtained as the corresponding
symmetric functions of the matrix
\be
\delta(a(\nu + g \varrho)/2) = \exp\left(-\ri a \sum_{k=1}^{n-1}
( \nu_k +g) \Lambda_k\right),
\ee
where $\varrho := \sum_{k=1}^{n-1} \Lambda_k$ and we remind the convention (\ref{I.6}).

The above very simple considerations reproduce the result of van
Diejen and Vinet \cite{vDV}, who determined the joint spectrum of the commuting Hamiltonians
${\cal H}_r^{\mathrm{op}}$  by means of Schr\"odinger quantization.
They proceeded by making
 the commuting  formal difference operators introduced previously by Ruijsenaars
self-adjoint on a finite-dimensional Hilbert space.
The Hamiltonians ${\cal H}_r^{\mathrm{op}}$ were then
diagonalized by a non-trivial calculation.
Their finite-dimensional Hilbert space is built on the lattice of the points
$\sum_{k=1}^{n-1} ( \nu_k +g) \Lambda_k$, with the condition in (\ref{redstate}),
viewed as a discretization of the global
particle-position variables.
Here we have explained
that both this discretization and the spectra of the Hamiltonians follow from
K\"ahler quantization. That is, K\"ahler quantization and Schr\"odinger quantization
give the same result.

Having reproduced the spectral result of \cite{vDV}, it would be interesting
to derive also their eigenfunctions and the associated
quantized duality map by geometric (quantization) methods.

\bigskip
\bigskip
\noindent{\bf Acknowledgements.}
C.K. would like to thank the  organizers of the conference CQIS (Protvino, January 2011) for the opportunity to present this work which
was supported in part
by the Hungarian
Scientific Research Fund (OTKA) under the grant K 77400.

\end{document}